# The potential of retrofitting existing coal power plants: a case study for operation with green iron

J. Janicka[1], P. Debiagi[2], A. Scholtissek[2], A. Dreizler[3], B. Epple[4], R. Pawellek[5], A. Maltsev[5], C. Hasse[2*]
*hasse@stfs.tu-darmstadt.de

[1] Technical University of Darmstadt, Department of Mechanical Engineering, Energy and Power Plant Technology
[2] Technical University of Darmstadt, Department of Mechanical Engineering, Simulation of Reactive Thermo-Fluid Systems
[3] Technical University of Darmstadt, Department of Mechanical Engineering, Reactive Flows and Diagnostics
[4] Technical University of Darmstadt, Department of Mechanical Engineering, Energy Systems and Technology
[5] STEAG Energy Services GmbH

**ABSTRACT**

Storing electrical energy for long periods and transporting it over long distances is an essential task of the necessary transition to a $CO_2$-free energy economy. An oxidation-reduction cycle based on iron and its oxides represents a very promising technology in this regard. The present work assesses the potential of converting an existing modern coal-fired power plant to operation with iron. For this purpose, a systematic retrofit study is carried out, employing a model that balances all material and energy fluxes in a state-of-the-art coal-fired power plant. Particular attention is given to components of the burner system and the system's heat exchanger. The analysis provides evidence that main components such as the steam generator and steam cycle can be reused with moderate modifications. Major modifications are related to the larger amounts of solids produced during iron combustion, for instance in the particle feeding and removal systems. Since the high particle densities and lower demand for auxiliary systems improve the heat transfer, the net efficiencies of iron operation can be one to two percentage points better than coal operation, depending on operating conditions. This new insight can significantly accelerate the introduction of this innovative technology by guiding future research and the development of the retrofit option.

## 1. MOTIVATION AND BACKGROUND

Supplying societies with sustainable, reliable and affordable energy is one of the most pressing challenges of our time. Combined with an energy crisis, the extreme weather events in recent years, from heavy rain to heat and drought, have drastically shown how urgent it is to convert from a fossil-based to a renewable energy system. This transition must address competitiveness and security of supply, as well as environmental and social sustainability. It has further become clear that the process of transforming our energy systems needs to accelerate. The current phase of the energy transition is characterized by increased electrification, accompanied by a significant increase in power generation from renewable sources. To compensate for the volatility of those renewables, new storage technologies are required. Transitioning to a renewable energy system will be the challenge of the century for humankind, as current energy systems depend heavily on fossil fuels. Figure 1 illustrates the world's primary energy supply (PES), the share of main energy sources and their contributions to total $CO_2$ emissions as of 2019. From a total PES of 168 PWh, natural gas, oil and coal combined still accounted for more than 80% (136 PWh) [1]. The consumption of fossil fuels generated a total of 34 Gt of $CO_2$ [2]. These plots highlight how coal is a major contributor to $CO_2$ emissions and why phasing out coal energy is a priority.

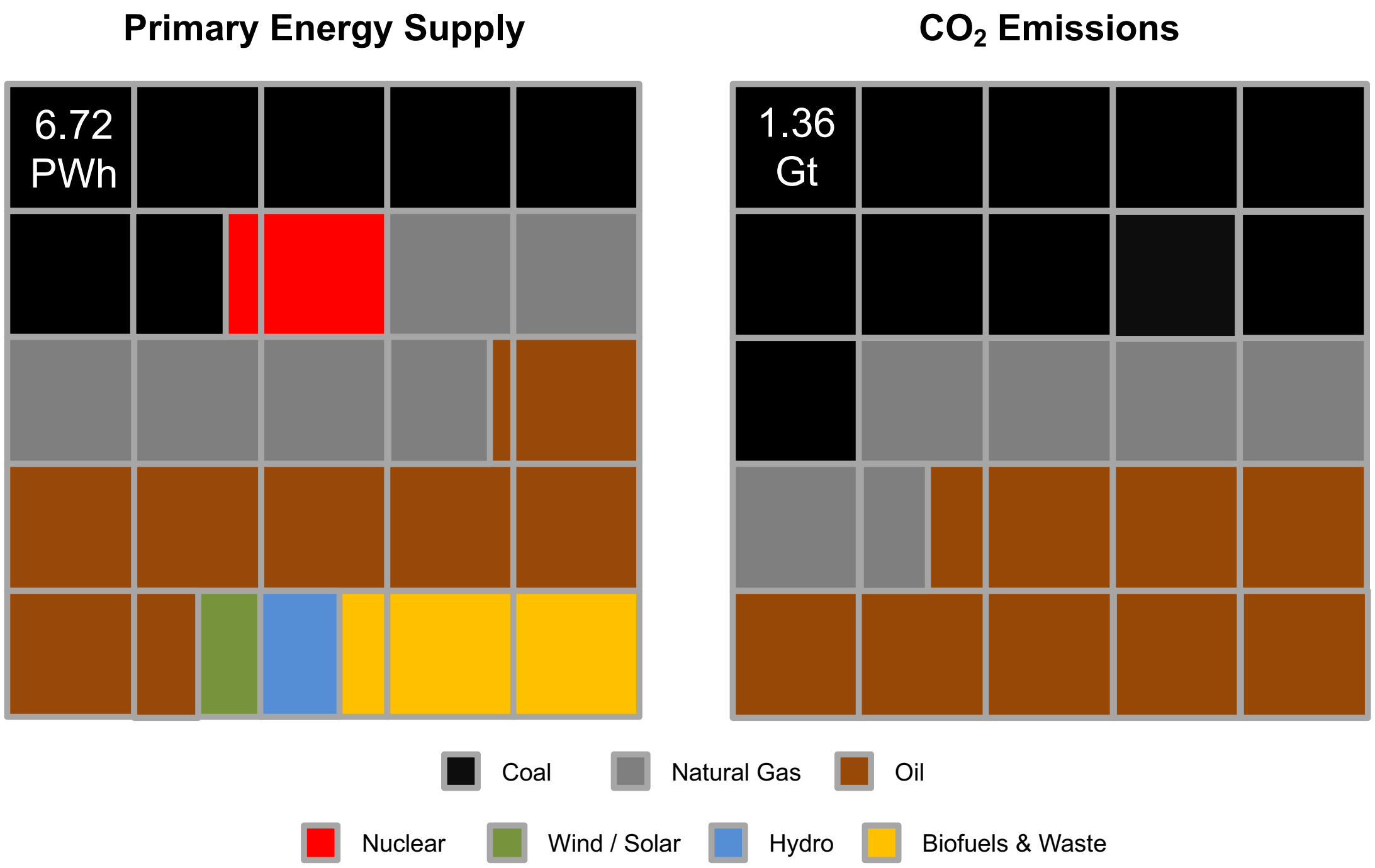

Figure 1 – World primary energy supply and associated $CO_2$ emissions [1,2].

Observing the global electricity sector, the scenario is slightly better: fossil fuels accounted for 61.3%, 35.1% being generated in coal power plants (9.4 PWh) in 2020 [1]. Local coal use varies, making up 60.75% in China, 23.66% in Germany and 13.22% in the EU-27 [3]. Currently (2022), 2449 coal power stations are operating worldwide, totaling 2.07 TW of capacity [4]. In order to reach the transition goal, the extensive electrification of all sectors is essential. The demand for electrical energy will increase significantly. Even if energy efficiency is increased by using electric power, about 40% of the secured and controllable capacity will have to be replaced by renewable energy sources, considering that nuclear and/or coal power is to be phased out in many countries [5,6]. In Germany, for example, 60 to 80 GW of on-demand capacity will be necessary when the country completes its plans to phase out coal and nuclear power plants. Therefore, the transition to renewable energy will not only require extensive investments to build new infrastructure; the current power generation capacity will also have to be raised due to the increased demand.

Wind power and photovoltaic plants have the largest capacity potentials for renewable electricity production. Windy and sunny areas that combine high expansion potential with great economic efficiency are typically located on the coasts, in the south of Europe or North Africa [7]. However, many sites of major energy consumption have limited potential for expansion, leading to a geographic separation between the production and consumption of renewable energies. Highly energy-demanding countries that have also limited renewable resources, such as Germany and Japan, will have to rely either on energy imported directly as electricity or on the long-distance trade of renewable energy carriers.

Moreover, the electricity generated from these plants is subject to strong production fluctuations (volatility), which additionally introduce a time offset between availability and demand. Short-term (day/night) fluctuations from domestic wind power and photovoltaic plants can be compensated for with technologies such as pumped-storage plants or batteries. In a 95% scenario[1], for example, major challenges arise such as: (1) long-term fluctuations resulting from cold, dark periods in winter, (2) the loss of secured power due to the replacement of nuclear and coal-fired power plants, and (3) the resulting need to transport renewable energies over long distances [5]. Therefore, without suitable storage, these renewable energies are not able to provide the necessary secured power plant capacity. Innovative, sustainable concepts for storing and transporting electricity generated from renewable sources are needed to overcome the spatial or temporal separation between the availability of and demand for electricity from renewable sources. This will require an unprecedented level of technological innovation, especially in the chemical energy carriers used to store renewable electricity. The direct use of hydrogen as an energy carrier, which has to be compressed or liquefied for transport, achieves a global energy efficiency of 11–31% [8,9] – this estimate is based on electrolysis efficiencies for hydrogen production between 50–74% and a thermochemical oxidation efficiency of 46% [10,11] which is comparable to electrification efficiencies of low-temperature proton exchange membrane (PEM) fuel cells [12], particularly at full load. Direct use of green hydrogen from remote partner countries, however, still requires technological improvements and development of appropriate transport capacity. As hydrogen is not easily stored or transported at utility scale, the storage of renewable energy in metals offers new opportunities to complement hydrogen technology.

[1] In the report "Impulses to shape the energy system up to 2050" [5], published by DENA, different energy transition scenarios are investigated. They particularly describe the future energy sector and the changes required for 80% and 95% reductions in greenhouse gas emissions.

The scientific background of this work is the utilization of metals as carbon-free energy carriers in an innovative energy cycle as a central building block of the energy transition (visualized in Figure 2). It introduces a decarbonized energy supply system with a secured power capacity based on a retrofitted power generation infrastructure, such as coal power plants operating with carbon-free metal fuels. This concept provides a solution for countries in which coal power plants are scheduled for retirement, and also for countries where there are no current alternatives for phasing out coal and the construction of modern coal power plants is already planned. In both cases the additional $CO_2$-free electricity produced could support energy transition plans. During the storage part of the cycle, electrical energy from renewable sources is chemically stored in metals by direct electrochemical reduction or through thermochemical reduction using green hydrogen as a reducing agent. When high-energy-density metals are used, the renewable energy can be stored and transported. On the release side of the cycle, the energy stored in reduced metals is released using different methods to obtain heat, power and/or hydrogen.

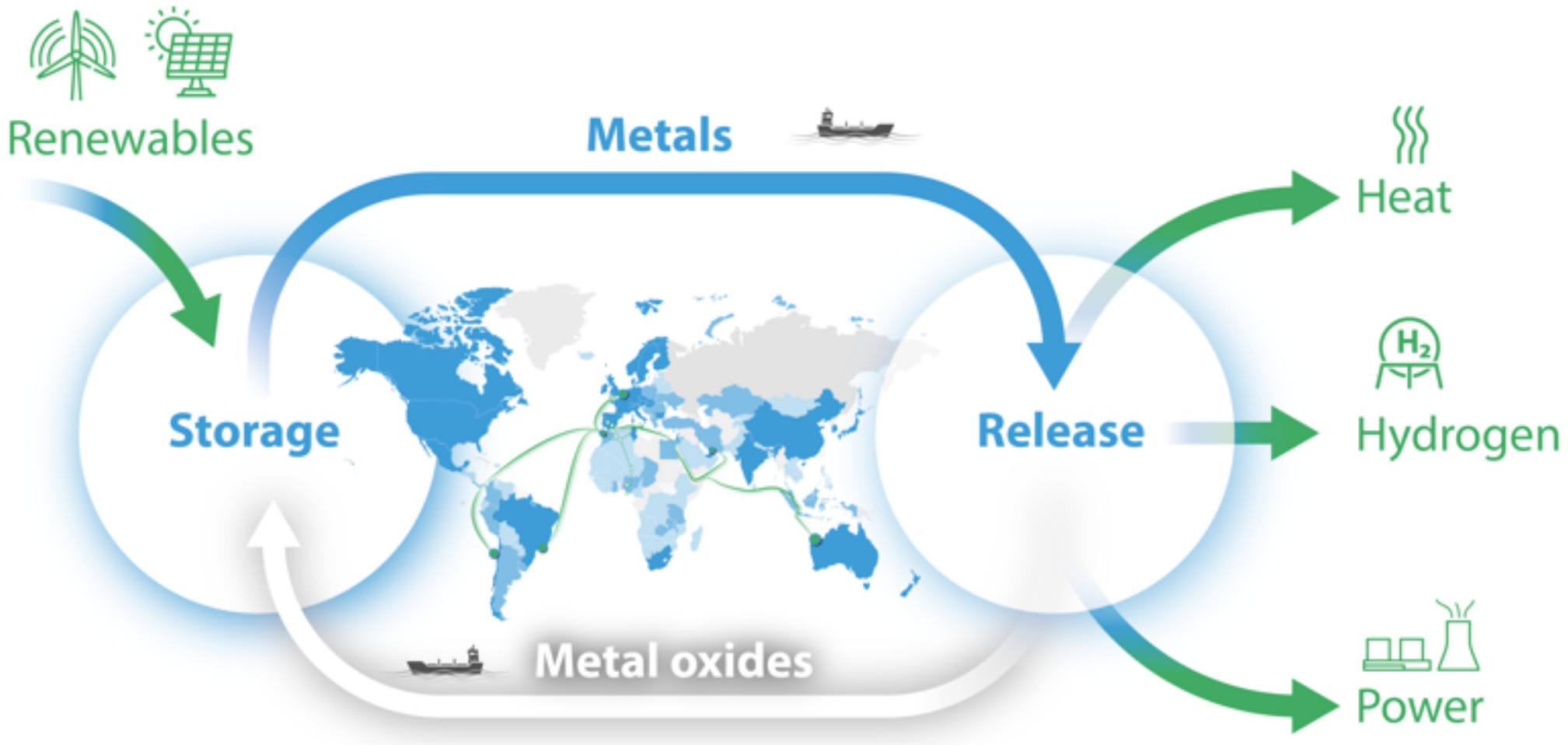

Figure 2 – The iron cycle: innovation for renewable energy storage.

The utilization of metals as energy carriers in this concept is extensively described by Bergthorson et al. [13,14], Trowell et al. [15], Yavor et al. [16,17], Julien and Bergthorson [18] and Schiemann et al. [19]. Dirven et al. [20] provided an assessment and compared coal with four metal fuels in terms of their overall efficiency on a complete cycle of power generation and fuel regeneration. Debiagi et al. [8] investigated the material and infrastructure demands for implementing this iron-fuel cycle, revealing that the current major bottleneck is not the availability of iron or the iron reduction capacity, but the limited installed capacity of renewable electricity and electrolyzers to produce Green $H_2$.

Among other high-energy-dense metal and semi-metal fuels such as aluminum, titanium and silicon, iron as a metal fuel is a very attractive solution for stationary power generation for several reasons: it is a stable and abundant material, it has a high volumetric energy density, low toxicity and low market price, and it boasts an existing production infrastructure and transportation network [8]. The high-temperature oxidation of Iron produces heat, similarly to the traditional combustion of solid fuels, but without releasing $CO_2$. The products of its combustion are iron oxides, which are solid under standard conditions and can easily be collected for recycling. Due to their properties, iron and iron oxides can be safely stored for long periods of time. As a result, using iron as a chemical energy carrier makes it possible to secure the supply of energy through controllable power plant capacities with an estimated energy efficiency between 18% and 29%, which is comparable or even superior to direct $H_2$ in some conditions [9]. This estimation is based on the average gross efficiency of a coal-fired power plant in 2018 [6], with potential for over 50% due to the special properties of iron, as discussed in this paper. A techno-economic assessment [9] of this concept, including the long-distance trade between potential renewable energy exporters and importers, highlights the higher sensitive and uncertain variables that need further development and projects the conditions to reach cost-competitive carbon-free electricity generation using iron in retrofitted coal power plants.

Traditional iron reduction for steel production is among the most $CO_2$-intensive activities, and the decarbonization of this industry goes hand in hand with using iron as a recyclable energy carrier, as scaling up such production processes is extremely beneficial to the industry in the long term [21]. Completely decarbonizing the steel industry could save 2.6 Gt of $CO_2$ emissions per year [21]. Retrofitting the world's active coal power plants for iron combustion would save approximately another 10 Gt of $CO_2$ emissions per year [2].

Against this background, several research groups have been working on developing a better understanding of high-temperature iron oxidation processes with a focus on the fundamentals of iron-air flames. Goroshin et al. [22] performed reduced-gravity experiments on laminar iron dust flames and Tang et al. [23] modeled the modes of combustion for iron dust particles using those experimental data. Hazenberg and van Oijen [24] developed an Eulerian-Lagrangian model to capture the flame structure and burning velocities of iron dust flames. Huang et al. [25] performed experiments to improve the understanding and prediction of micro-explosion events in pulverized iron combustion. Ning et al. [26] experimentally measured the time-resolved temperature for iron particles using laser ignition, and characterized several properties of the post-combustion particles. Li et al. [27] investigated particle melting and nanoparticle formation. In summary, aspects such as flame velocities, ignition and flammability behavior, and undesirable aspects such as nanoparticle formation, have already been investigated [24,26,28–30], revealing that the combustion properties of fuel gases such as methane and iron dusts with particle sizes of 10 to 20 µm are comparable [28].

The attractiveness of introducing a cycle of this kind could be significantly increased if it were possible to continue using as many components of the existing infrastructure as possible. This would save considerable investment costs, shorten implementation times and extend the life cycle of hundreds of power plants scheduled for decommissioning. This work is intended to contribute to this task and pursues the following objectives:

- Analyzing the potential of existing coal-fired power plants for retrofitting with iron as a fuel;
- Identifying the essential retrofitting measures for the power plant;
- Identifying the central unanswered scientific and technological questions about the retrofitting process.

## 2. POWER PLANT CHARACTERISTICS AND SIMULATION MODEL

To meet the above objectives, a case study for retrofitting an existing coal-fired power plant is carried out. A highly detailed and sophisticated model for the coal-fired power plant is extended to include iron operation. The power plant under consideration is a modern coal-fired power plant of the 800 MW class, and went into operation about 10 years ago. The main technical characteristics are summarized as follows:

1) Steam parameters are 600 °C and 620 °C for the high-pressure (260 bar) and intermediate-pressure sections (65 bar), respectively;
2) The preheating process is realized by five high-pressure preheaters and three low-pressure preheaters separated by a feedwater tank;
3) The condenser water is cooled by a wet cooling tower;
4) The turbine set is composed of single-flow high- and medium-pressure turbines and a double-flow low-pressure turbine.

The operating conditions with the essential thermodynamic data are presented in detail in the results section of this paper. A comprehensive simulation model has been created for this power plant with the software EBSILON® Professional [31], which represents the process in all essential components and also allows transient processes to be studied [32]. Figure 3, showing the water-steam part of the power plant, gives an impression of the level of detail of the model.

The simulation model consists of more than 1000 components and contains all essential control loops. All central components such as heat exchangers, turbines and pumps have been adapted to real operating conditions via characteristic curves or approximation polynomials. For the heat exchangers, a so-called efficiency function is used, representing the ratio of the calculated heat transfer coefficient to the measured heat transfer coefficient. For the turbines, a ratio of measured to calculated efficiencies is used. All calibration information is then stored as a large database of operating parameters using approximation functions. This procedure allows the model to almost exactly reproduce the measured operating conditions. EBSILON® Professional is used for three main applications by the operators of the power plant:

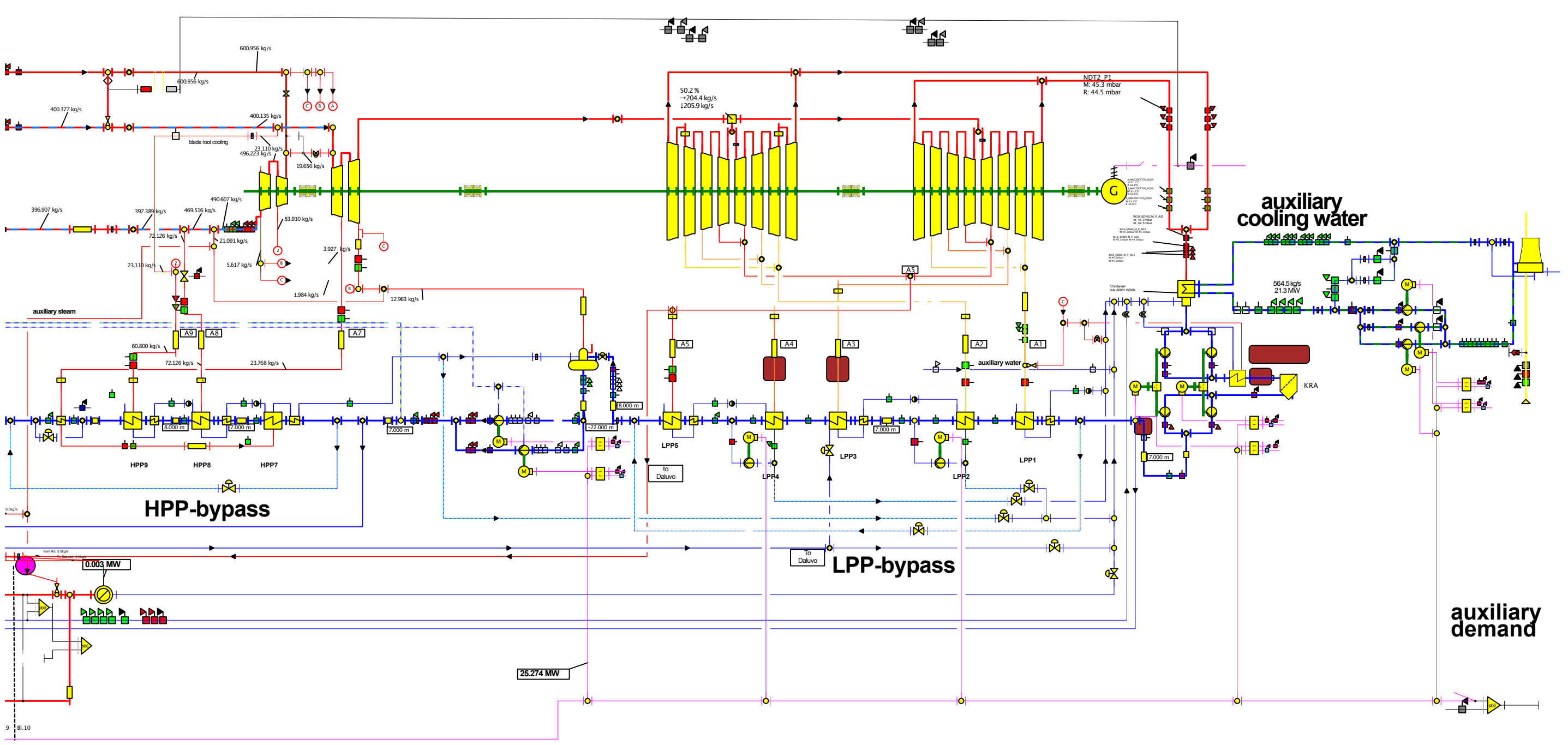

Figure 3 – Water-steam cycle of the power plant under consideration. Full-size figure is available in the supplementary material.

**Process identification and monitoring:** While the plant is in operation, so-called identification simulations are carried out at five-minute intervals. These involve comparing the stored efficiency function and efficiencies with the measured values. This allows conclusions to be drawn about deteriorations in the plant's condition. Thus, possible deterioration can be identified for every plant component.

**"What-if" calculations:** In these calculations, again based on the correction curves calibrated with real operating conditions, future scenarios are calculated in advance.

**Measurement data reconciliation:** Due to the high redundancy of measuring points, predictions can be made via the integrated algorithms as to whether the respective measuring devices are still functional within their tolerance range or require maintenance/replacement.

Another impression of the model depth is given by Figure 4, which shows the boiler components. The furnace is divided into a total of nine flue gas heat exchangers. The first four elements describe the furnace. The superheater is modeled by three elements and the reheater sections by two elements. One segment represents the feedwater preheater. The heating surfaces on the steam side are divided into nine main heating surfaces and thirteen secondary heating surfaces such as support tubes or wall heating surfaces.

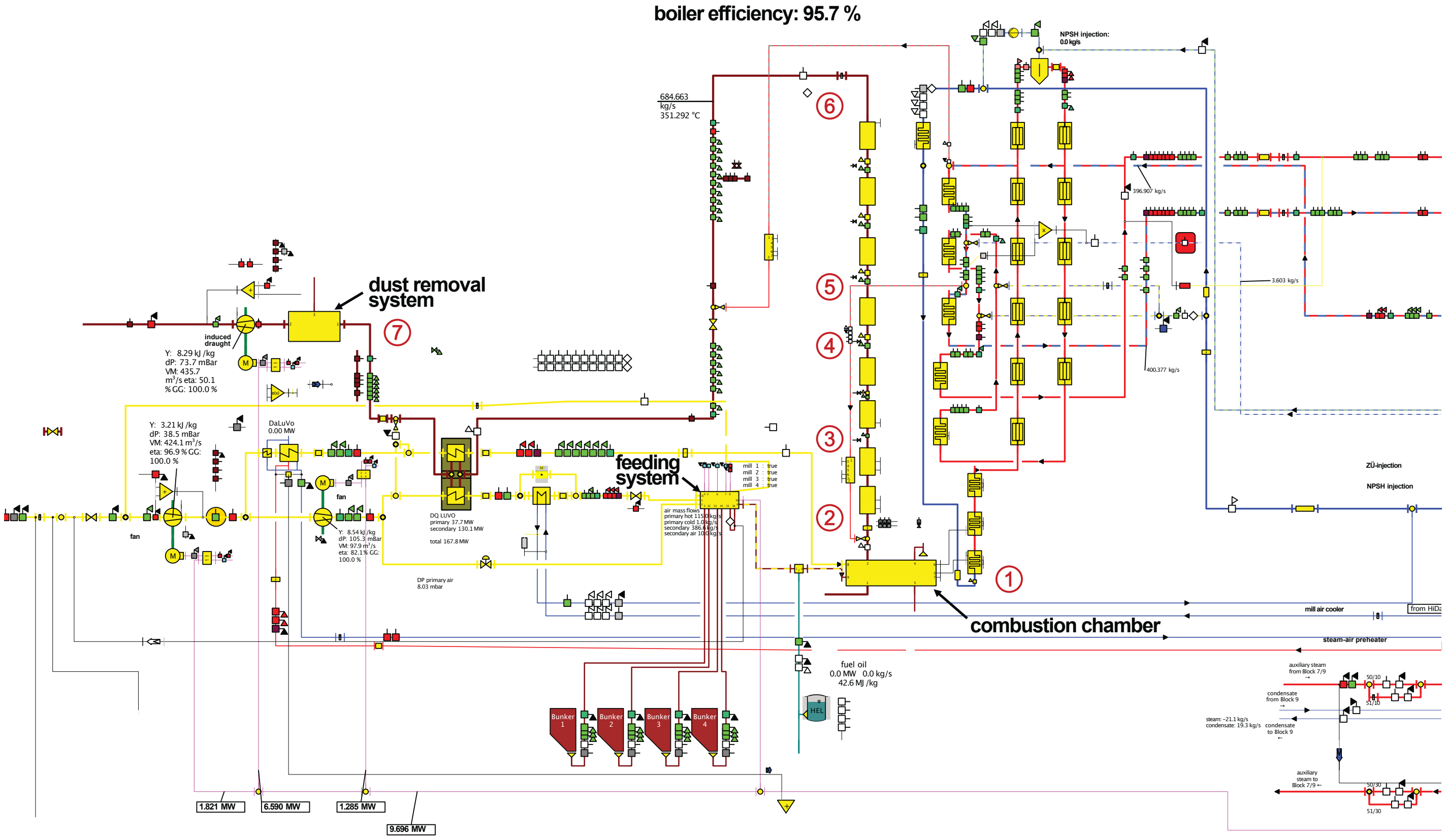

Figure 4 – Boiler and flue gas-air system of the power plant under consideration. Full-size figure is available in the supplementary material.

This detailed simulation model, which has been calibrated to coal-firing operating conditions, was substantially extended to simulate different scenarios with iron as the fuel. Figure 5 shows a simplified schematic of the overall process highlighting the primary parts of the power plant, the combustion system and the steam cycle. In the combustion section, the iron fuel is supplied by a feeding system along with air that is transported by a fan to the combustor. After oxidation, the iron oxides are collected and the remaining exhaust gas is removed by a second fan. The combustion process is connected to the steam cycle by heat exchangers that transfer the heat of combustion, $\dot{Q}_B$, to the boiler. The most important parts of the steam cycle are the steam turbine producing the electrical power, $P_e$, the condenser, and two preheating systems (high and low pressure, respectively). Both parts of the process rely on auxiliary power for operation. The energy and power streams, as indicated in the schematic, are used to define the efficiencies for the iron oxidation process (cf. Table 1). The boiler efficiency results from the ratio of the heat flow supplied to the cycle and the heat flow supplied to the boiler in line with DIN EN 12952 [33]. This is essentially determined by the calorific value of the fuel. The cycle efficiency is the ratio of the electrical power output at the generator to the heat flow supplied by the cycle. The auxiliary efficiency is calculated from the output power at the generator, minus the energy demand of all plant components, divided by the output power of the generator. Multiplying these three efficiencies then gives the total or net efficiency of the power plant.

Table 1 – Efficiencies of the iron oxidation process. Energy and power streams relate to the schematic shown in Figure 5.

| | |
|---|---|
| Total process efficiency | $\eta_t = \frac{P_e - \sum_i P_{a,i}}{P_{\text{fuel}}}$ |
| Cycle efficiency | $\eta_c = \frac{P_e}{\dot{Q}_B}$ |
| Efficiency with respect to auxiliary power | $\eta_a = \frac{P_e - \sum_i P_{a,i}}{P_e}$ |
| Boiler efficiency | $\eta_B = \frac{\dot{Q}_B}{P_{\text{fuel}}}$ |

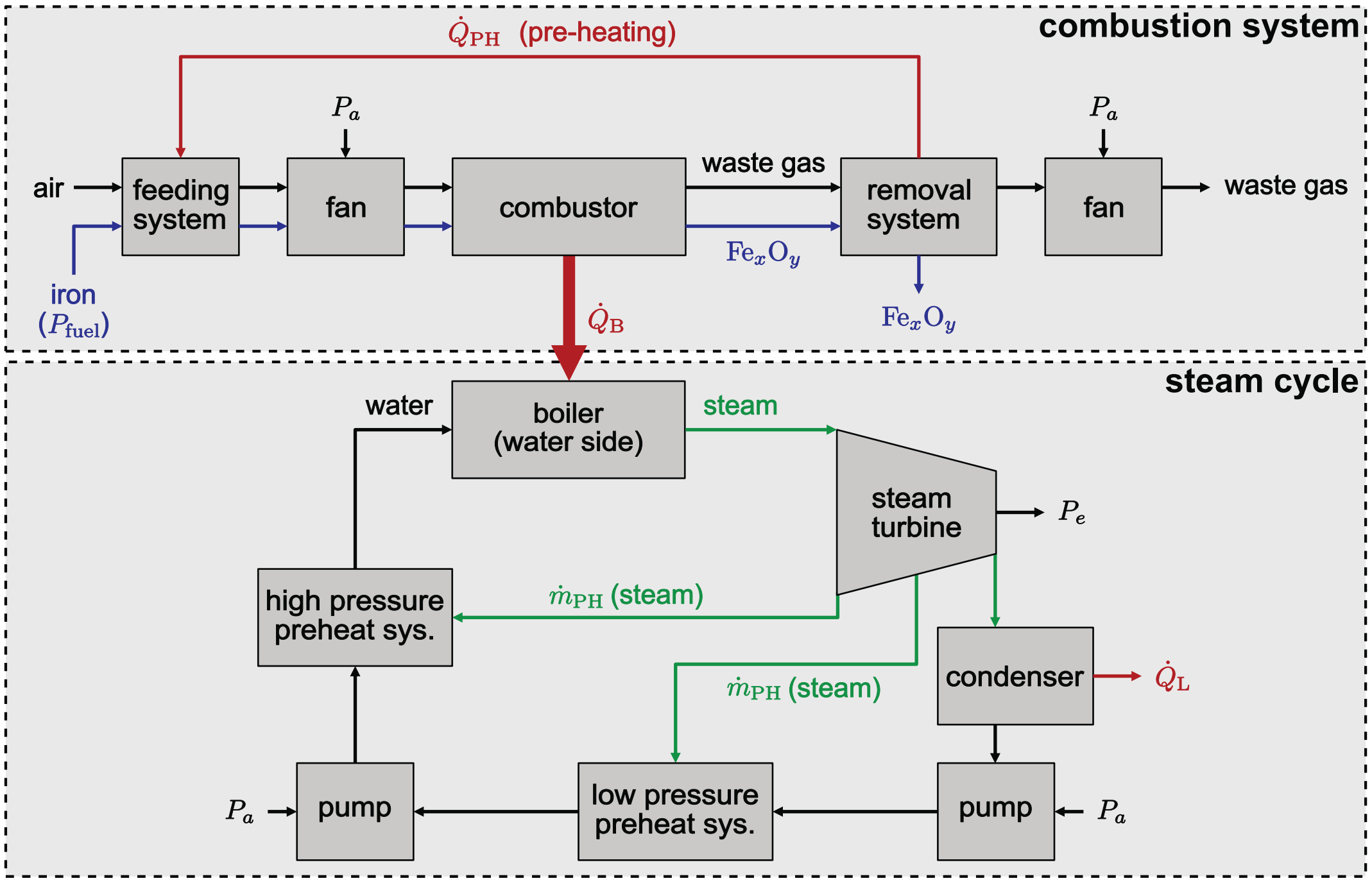

Figure 5 – Simplified schematic of the retrofitted power plant. The process can be divided into a combustion system and a steam cycle, which are connected via heat exchangers between the combustor and the boiler.

## 3. THERMODYNAMIC CONSIDERATIONS OF THE HIGH-TEMPERATURE IRON OXIDATION

Understanding the physico-chemical differences between coal and iron as fuels, as reported in Table 2, is essential to investigate the modifications required in a power plant. Bituminous, sub-bituminous and lignite coal types are typically used in power plants, and they mostly differ in terms of their energy density. Iron and iron oxides are highly dense materials. The heating value in the oxides decreases progressively depending on the oxidation state, reaching zero in hematite form ($Fe_2O_3$). Compared with typical bituminous coals, iron exhibits a much higher density, but a lower gravimetric heating value. However, the volumetric energy density of iron is significantly higher, which is advantageous, since the volume occupying transportation and storage units and the volumetric flow in burners would be smaller than those of coal for the same energy content.

Table 2 – Properties of different coal ranks, iron and iron oxides. Adapted from [8]

| Fuel | Density [kg m$^{-3}$] | Melting Point [°C] | LHV [MJ kg$^{-1}$] | LHV [GJ m$^{-3}$] |
|---|---|---|---|---|
| Fe | 7870 | 1538 | 7.36 | 57.92 |
| FeO | 5740 | 1377 | 2.03 | 11.65 |
| $Fe_3O_4$ | 5180 | 1590 | 0.35 | 1.81 |
| $Fe_2O_3$ | 5260 | 1565 | 0 | 0 |
| Bituminous Coal | 1200–1600 | - | 25–36 | 30–57 |
| Sub-Bit. Coal | 900–1200 | - | 20–25 | 18–30 |
| Lignite | 700–900 | - | 10–20 | 7–18 |

In comparison to coal, the high-temperature oxidation of iron is more complex, as can be seen from Figure 6, which is adapted from reference [34]. When the $O_2$ contents in the exhaust gas are those of an iron combustion system, i.e., between two and eight percent with air factors between 1.2 and 2.0, the green shaded area in Figure 6 is obtained. Above the melting point of $Fe_3O_4$ (magnetite, $T_{melt}$ = 1597 °C) a liquid iron oxide phase is observed. In an intermediate range between this temperature and about 1300 °C, crystalline $Fe_3O_4$ is found. Below this temperature, iron oxide is present as crystalline $Fe_2O_3$ (hematite). Figure 6 shows further simulation results obtained with the equilibrium program developed by NASA [35], which is part of the simulation software EBSILON® Professional. Although the equilibrium calculation is based on ideal liquids and ideal gases without a mixing model, very good agreement is observed in the regime of interest for the high-temperature oxidation of iron. At $O_2$ concentrations below $10^{-5}$ atm, however, which are not of interest for the present analysis, the discrepancies become much larger due to the ideality assumption. For example, between $10^{-5}$ atm and $10^{-10}$ atm, no liquid phase is obtained in the equilibrium calculations.

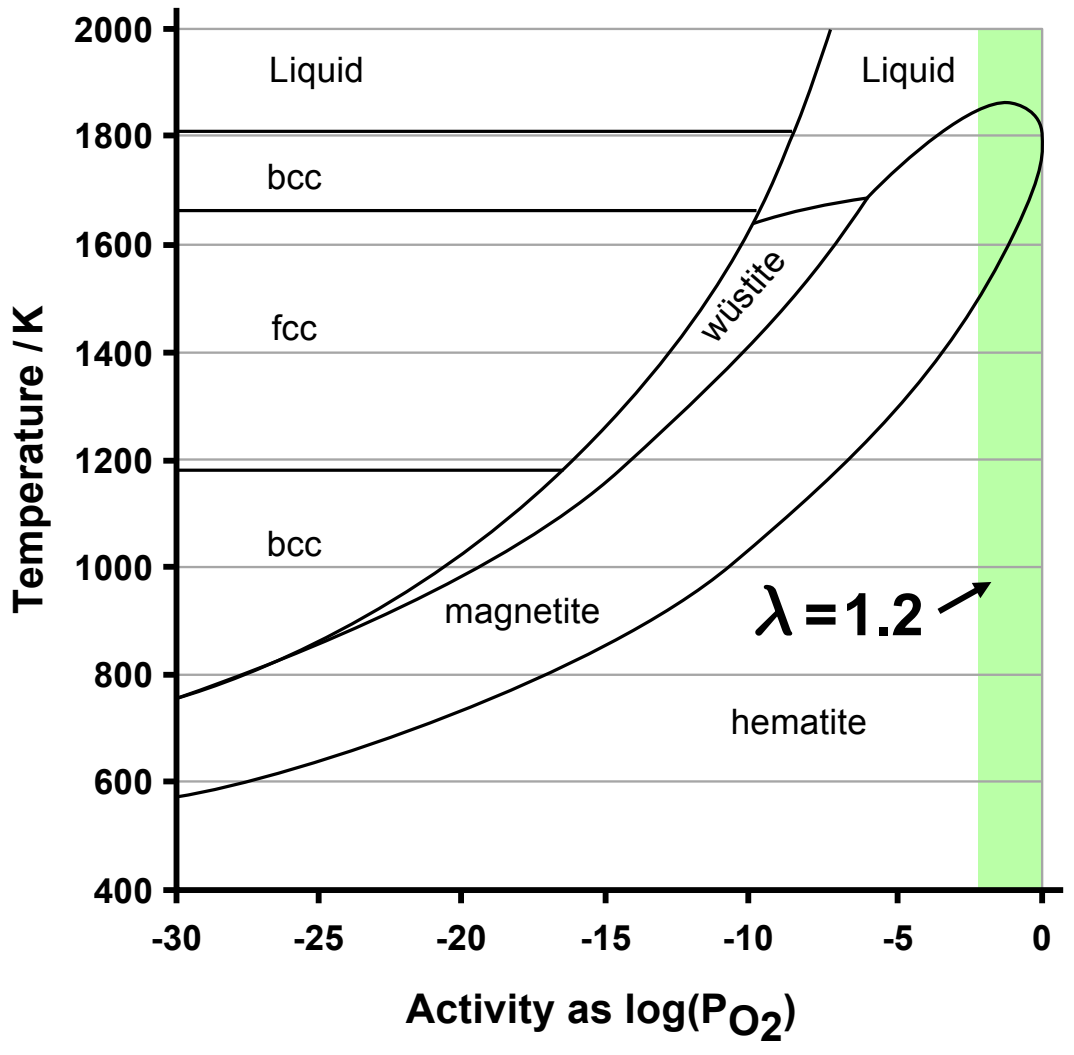

Figure 6 – Phase diagram of the Fe-O system as a function of temperature and $O_2$ partial pressure (adapted from [34]).

If iron is oxidized in the crystallization state at adiabatic conditions with an air factor of $\lambda$ = 1.2 at an initial temperature of 20 °C, an adiabatic oxidation temperature of 1748.7 °C is obtained if it is further assumed that the liquid phase consists of pure $Fe_3O_4$. Given the same assumptions, reference calculations with another equilibrium solver (OpenSMOKE++ [36]) verify these results (deviations below 0.4 °C). When the mixture of oxidation gases and $Fe_3O_4$ is cooled and an equilibrium composition is assumed at the respective temperature, a heat release of 7380 kJ/kg, corresponding to the heating value of iron, is obtained. Figure 7 shows the heat-temperature diagram for this cooling process per kilogram of iron.

Starting from the adiabatic combustion temperature, liquid $Fe_3O_4$ is first cooled. At the melting temperature of 1597 °C, the heat of melting, which is about 10% of the calorific value, is released. Then $Fe_3O_4$ is cooled to a temperature of about 1300 °C. At this temperature, $Fe_3O_4$ transforms into $Fe_2O_3$ with the further release of heat, which again is about 10% of the heating value. When this crystalline phase is cooled to the reference state at 20 °C, the total heat corresponding to the calorific value has been released.

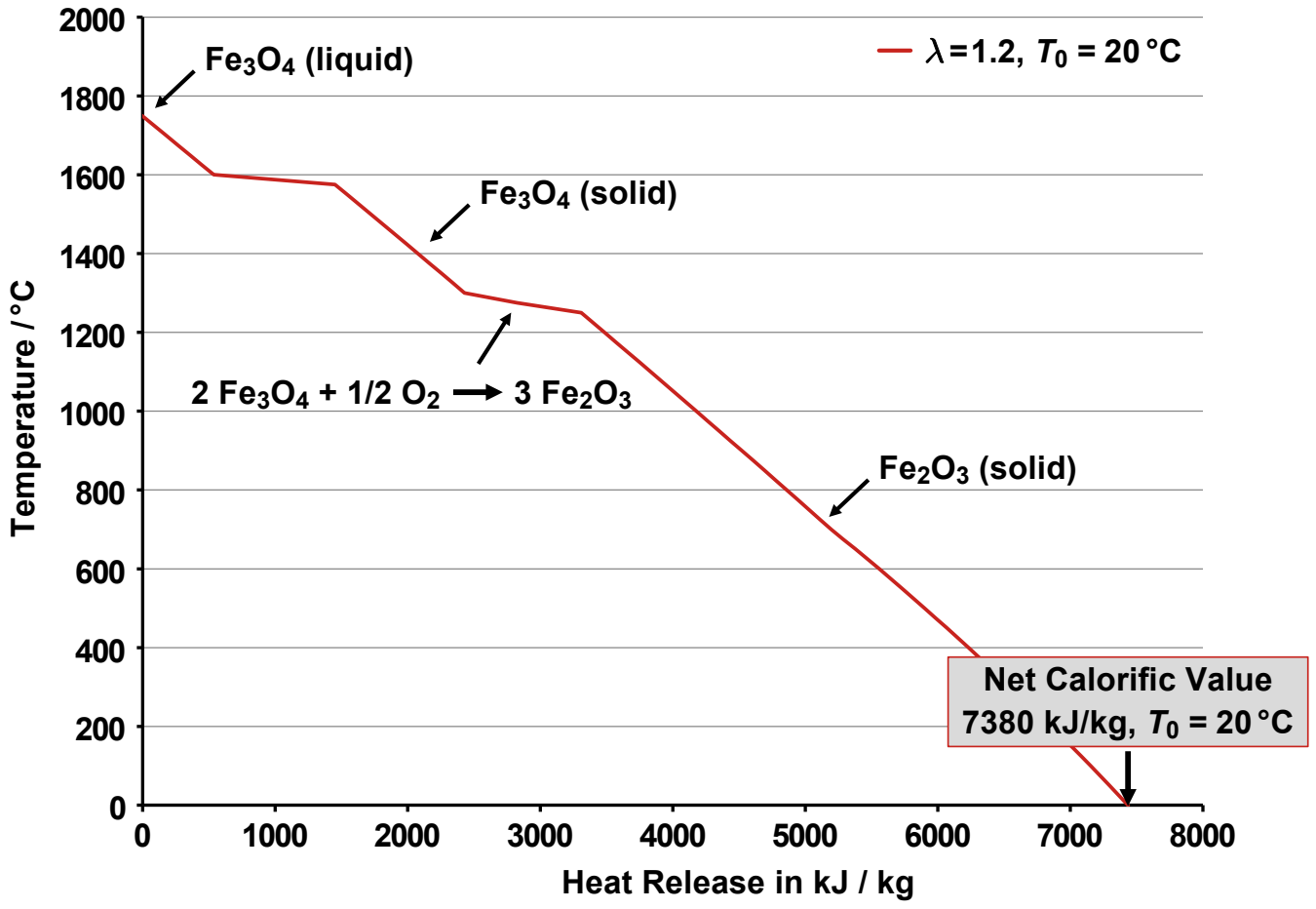

Figure 7 – Heat release of iron, which can be obtained when the material is cooled after oxidation to ambient conditions (here: $T_0$ = 20°C).

In power plants, however, the oxidation air is preheated via air preheaters with hot exhaust gases, such that temperatures between 250 and 300 °C are reached after the fuel is fed upstream of the combustion chamber. If an average preheating temperature of 275 °C is taken as a basis, the adiabatic combustion temperature rises to 1940 °C. For hard coal with a calorific value of about 25000 kJ/kg, which is otherwise used in the power plant simulation, by coincidence almost the same adiabatic combustion temperature is obtained. Figure 8 shows the cooling curves normalized to the respective calorific value for both fuels.

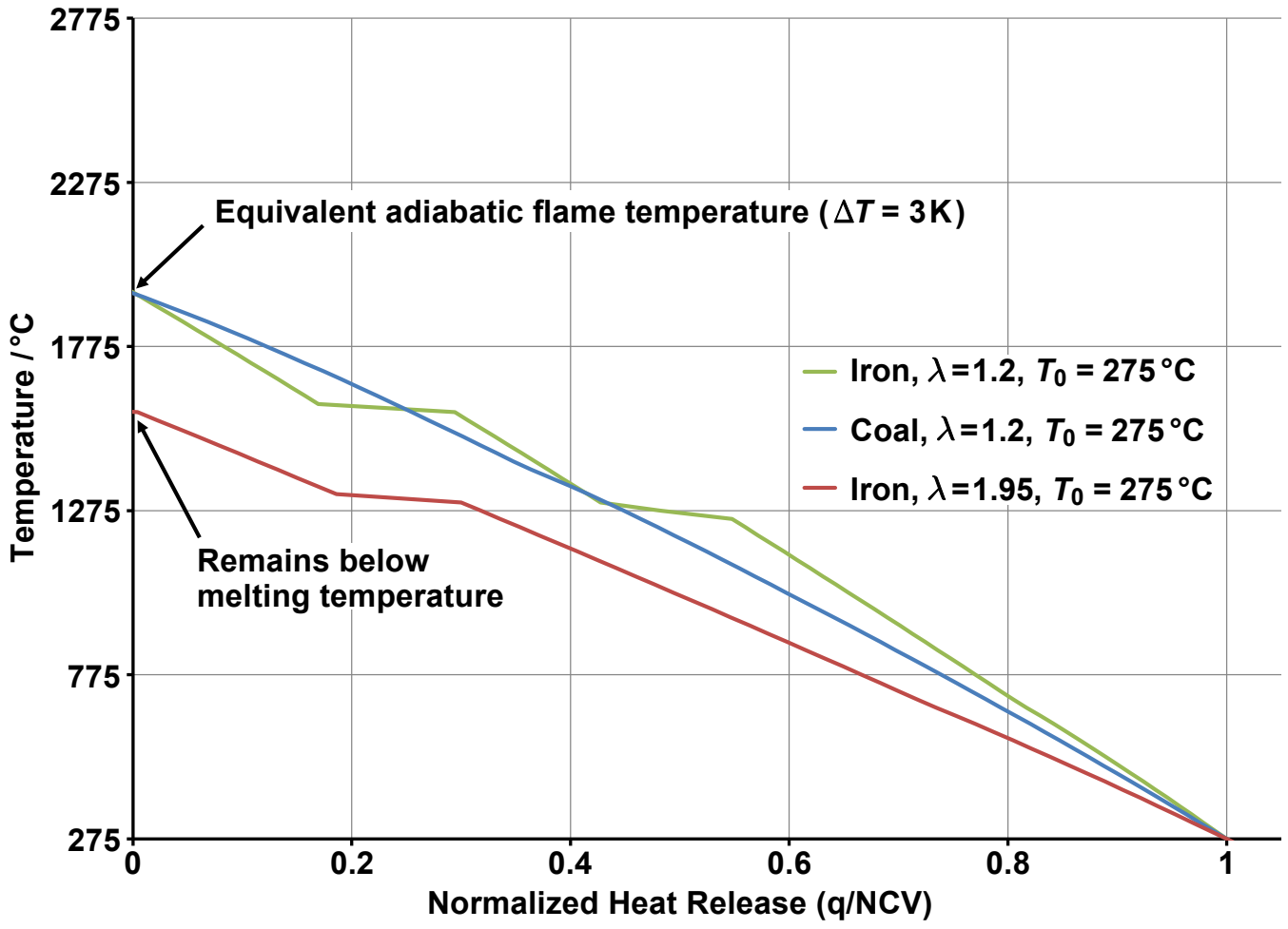

Figure 8 – Normalized Q-T diagram for iron at different air factors and preheating temperatures of 275 °C.

Furthermore, Figure 8 shows the heat temperature curve for an oxidation process with an air factor of $\lambda$ = 1.95, which, assuming a perfect homogeneous mixture, has an adiabatic combustion temperature that is just below the melting point of $Fe_3O_4$ at 1596 °C. Thus, a lower particle agglomeration effect and lower heating surface fouling can be expected. This aspect will be discussed in more detail below.

## 4. TECHNICAL BOUNDARY CONDITIONS AND NECESSARY RETROFIT MEASURES

In addition to the thermochemical properties, the particle size distribution is of decisive importance for the conceptual design of the iron oxidation process. It is immediately apparent that iron grinding processes would be very costly and significantly reduce the efficiency of the overall process. In the subsequent considerations regarding the design of the iron oxidation process, it is assumed that the particle size distribution is adjusted once, for example by atomizing liquid iron, and that it does not change significantly during recurrent oxidation and reduction process steps. Therefore, it is necessary to avoid particle agglomeration. There is a high chance of achieving this if only crystalline particles are present. Liquid particles, in contrast, tend to agglomerate. Hence, the oxidation (and reduction) processes are ideally designed such that the melting temperature is not exceeded and additional measures such as exhaust gas recirculation or fuel staging could help mitigate particle agglomeration. Otherwise, it would be necessary to regenerate the iron particles by melting and atomization after every cycle. This would entail additional efficiency losses of around 12–18% of the reduction process step, depending on the temperature after reduction.

For the global analysis of the conditions for retrofitting a coal-fired power plant for operation with iron, it is important to examine the main operating conditions. Table 1 shows the conditions of the above power plant with an air factor of $\lambda$ = 1.17—the underlying reference case—and two conditions for iron oxidation. For better comparability, iron simulations were also performed with this air factor of $\lambda$ = 1.17. As can be expected, in these conditions a liquid phase of $Fe_3O_4$ is present, which could lead to particle agglomeration, as discussed above, and the deposition of the material on the boiler walls. However, it should be noted that the estimates in this work are based on thermodynamic equilibrium calculations and therefore cannot take into account finite rate chemistry effects. Due to kinetic effects, the peak temperature will be lower. Furthermore, internal staging of fuel or air is conceivable, which also has a temperature-reducing effect. It is also unknown whether collisions between liquid particles under operating conditions result in agglomeration or in elastic collisions between the particles. For reference, a second case with an air factor of $\lambda$ = 1.95 is considered, which safely ensures average temperatures below the melting temperature.

Table 3 reports that the mass fluxes between operation with coal and iron do not differ greatly for the same air factors, but the cases with iron exhibit a significantly higher solids content before the reaction (22–33 wt.%) and especially after the reaction (33–45 wt.%). This high particle concentration after the reaction is the decisive difference to coal combustion. The particle concentration before the reaction is also significantly higher than that of the hard coal-fired plant considered here, but is in the same order of magnitude as for lignite-fired power plants before the drying process.

Table 3 – Operating conditions for coal- and iron-firing.

| | Units | Coal operation with air factor $\lambda = 1.17$ | Iron operation with air factor $\lambda = 1.17$ | Iron operation with air factor $\lambda = 1.95$ |
|---|---|---|---|---|
| **Mass flux** | kg/s | 742 | 671 | 1008 |
| **Solids content (*prior* to reaction)** | wt.% | 9 | 33 | 22 |
| **Solids content (*after* reaction)** | wt.% | 1 | 45 | 31 |
| **Exhaust gas velocity** | m/s | 11 | 5.8 | 10.8 |
| **Density** | kg/m³ | 1800 | 8000/5200 | 8000/5200 |
| **Mean particle diameter** | µm | 80 | 15 | 15 |
| **Heating value** | kJ/kg | 25000 | 7380 | 7380 |

The choice of the iron particle size is of crucial importance for the operation of the power plant with iron. The analysis is based on two considerations: First, reaction kinetic aspects are important. The iron particles must be converted to iron oxide in the form of $Fe_2O_3$ or $Fe_3O_4$ in the furnace. Initial estimates documented in [28] suggest grain sizes between 10 and 20 µm. Second, one criterion when determining the iron particle size is whether unburned or partially burned iron particles can potentially be deposited in the ash hopper. This has to be avoided. Excessively high deposition values seriously worsen the efficiency. This is a more significant aspect for iron power plants compared to coal-fired power plants because, on the one hand, the sinking velocity of iron is greater due to its higher density, and on the other hand, the flow velocities are lower (see Table 3). The sinking velocity is determined based on the Reynolds number [34] and on temperature- and concentration-dependent material properties.

Figure 9 visualizes this relationship. It shows the ratio between the sink rate of the particles and the gas flow rate in the iron-fired power plant compared to the coal-fired power plant. When this ratio is required to be the same, the maximum particle size of 200 µm and mean particle size of 80 µm in the coal-fired operation correspond to particle sizes for the iron-fired operation of 40 µm and 15 µm, respectively. Thus, the particle size distribution obtained from this criterion is comparable to the findings from the reaction kinetic analysis [28].

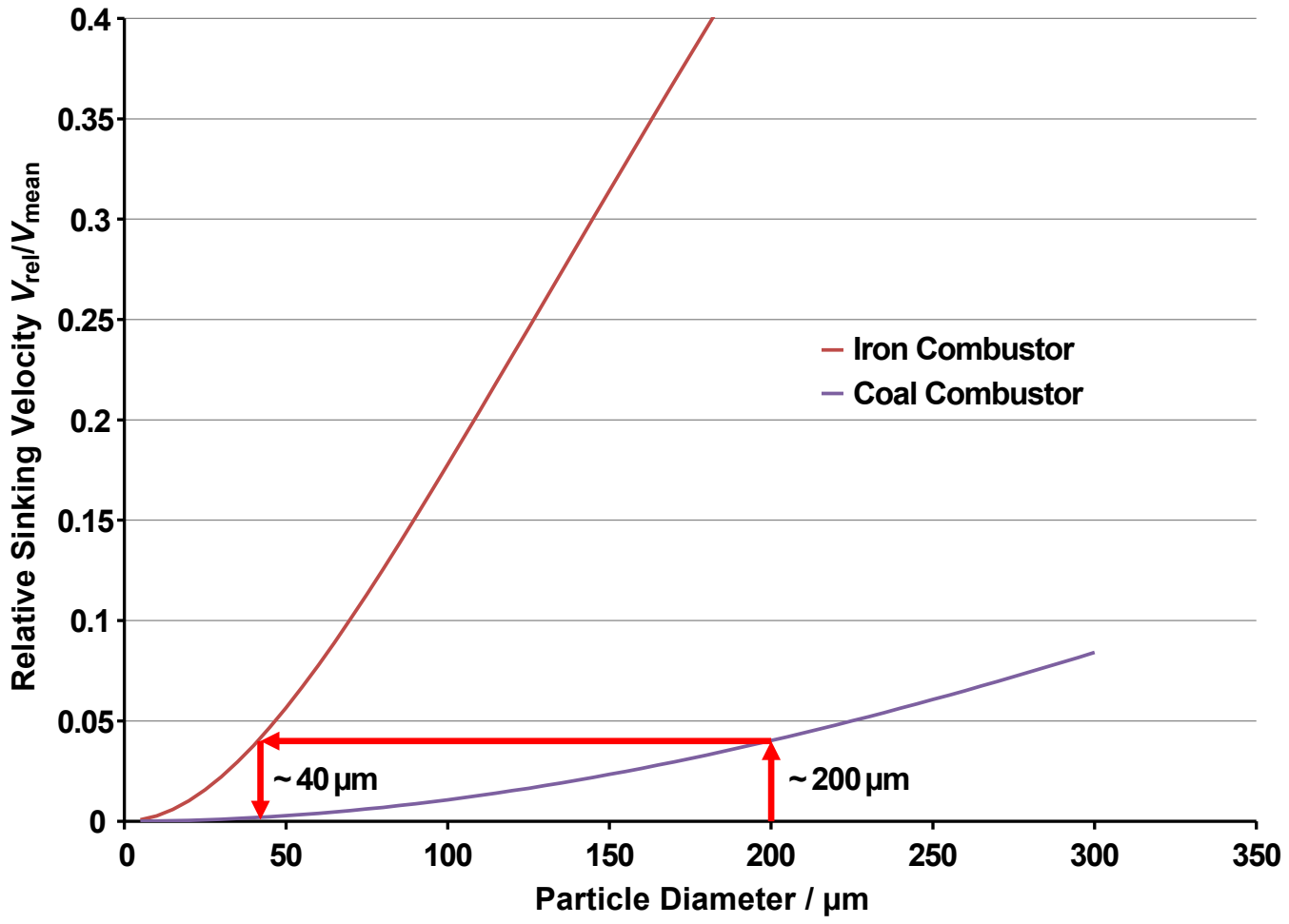

Figure 9 – Relative sinking velocity $V_s/V_{mean}$ as a function of particle diameter for the coal and iron operations.

## 5. MODEL EXTENSIONS FOR THE IRON RETROFIT ASSESSMENT

Retrofitting an existing coal-fired power plant for operation with iron requires the following plant components to be adapted:

**Fuel feeding system:** The feeding system must be converted for operation with iron. Mills are not necessary because, as previously explained, the particles are already delivered in a suitable particle size distribution. The mass flow of solids is higher than for a hard coal-fired power plant, but it is of the same order of magnitude as for lignite-fired power plants. Designing a feeding system of this kind is a classical engineering task.

**Solid fuel burners**: The burners must be redesigned to meet the conditions for iron operation. Again, based on the solid mass flows, it is easier to modify an existing lignite-fired power plant than an existing hard-coal-fired power plant because the jet burners should require less modification. With swirl burners, there is an increased risk that the largest particles will be deposited in the ash hopper.

**Flue gas dedusting**: In contrast to coal firing, the high dust contents of iron oxide must be separated from the exhaust gases. At present, it is not clear whether efficient electric filters can be used for this task since the ionization properties of the iron oxides have not yet been adequately investigated. For the present simulations, cloth filters were used as a basis. Initial estimates show that 10 to 12 filter units connected in parallel and two units in series can fulfill the separation task and satisfy particle emission limits. The relatively high pressure drop associated with these devices (assumption: 4000 Pa) was taken into account in the simulation [37].

**Removal of redundant components**: Mills and the desulfurization unit are not required, hence, restrictions such as sulfur dew point undershoots in the flue gas segment or compliance with maximum temperatures at the mills in the air segment do not have to be satisfied by the design of the air-flue gas system. Furthermore, since no $NO_x$ is formed via the fuel and the maximum temperatures are low thanks to high premixing to avoid melting, it is highly probable that – similarly to modern gas turbines – no additional secondary denitrification technology will be required.

**Air preheater:** It is not yet clear whether the rotary air preheaters used in coal power plants can also be operated with iron without major modification. Due to the relatively fine channels and the low wall thicknesses of the material, there is a risk of increased wear due to contact with the solid particles and an increased risk of clogging. In the simulations, it is assumed that these heat exchangers can be used. The authors are convinced that the use of a rotary heat exchanger for iron operation is not a fundamental problem. However, the measures required to clean and possibly protect the heating surfaces must be elucidated in operating tests.

**Cleaning and protection systems**: Due to the high dust concentration, cleaning systems must be provided that are operated with compressed air or steam to remove dust deposits from heating surfaces. Since there are no liquid particles in most of the boiler, the cleaning tasks can be solved using classical design methods found in mechanical process engineering. Furthermore, it may be necessary to protect the first rows of tubes of the convection heating surfaces from erosion by means of half-shell tubes.

**Biflux heat exchanger:** As will be explained later, shifts in heat quantities between the high-pressure section and the intermediate-pressure section of the boiler are to be expected in a retrofit depending on the air factor and the heat transfer conditions. In this study, a Biflux heat exchanger is used between the heating surfaces of the superheater. The simulation shows that there is no deterioration in the efficiency of the water vapor circuit. In a specific retrofit situation, however, it should be considered whether changes can be made to the piping instead.

## 6. SIMULATION OF IRON OPERATING CONDITIONS

Based on these preliminary considerations, a systematic simulation study is performed, varying different operating parameters. The following conditions are used:

**Steam cycle:** The entire water vapor system and all heat exchanger surfaces remained unchanged in comparison with coal operation.

**Heat transfer adjustments:** The VDI algorithms [38] for calculating the heat transfer, the α-values for convection and radiation on the flue gas and water vapor side and the material properties of the walls also remained unchanged. One exception was the calculation of particle radiation due to the high particle density, which requires backward scattering to be taken into account. Here, the VDI correlations were used for this specification [38]. During coal operation, this extension is unnecessary due to the lower concentration of solids. As will be explained later, the selected correlations are conservative estimates, especially for convective heat transfer.

**Efficiency factors:** The efficiency functions defined above, describing the relationship between the measured and calculated heat transfer coefficients, were also taken from operation with coal. These coefficients therefore include deviations between the idealized models (VDI algorithms) and the real design of the plant. These efficiency factors and adjustments include spatially averaged inlet and outlet temperatures and neglected inflow effects, or take into account the fouling of heating surfaces due to deposits of slag and ash. This adoption of the efficiency factors used in coal operations is also a conservative estimate, since the high melting points of $Fe_3O_4$ at 1597 °C presumably result in lower fouling in the radiation section of the boiler compared to coal ashes, which have significantly lower melting points (between 1000 and 1300 °C).

**Process losses:** Similarly, loss estimates were taken on unchanged from the coal simulation. These include radiation losses in line with DIN EN 12952 [33] of about 0.25%, mass losses of about 0.1%, and losses due to unburned fuel of about 0.5%. The same applies to all false air flows[2], mass flows for blade cooling, and machine efficiencies, as well as condenser water flows.

**Pressure constraints:** All restrictions, such as the minimum values for high-pressure and intermediate-pressure injection or the minimum values for the temperature levels of preheaters and aftercoolers, were set identically to coal operation. Pressure drop calculations were performed using the same algorithms as for coal. In addition, as already mentioned, pressure losses due to dust removal in the cloth filter were taken into account.

**Model extensions:** New simulation models were developed for the feed system, the burners, the dust removal and the Biflux heat exchanger.

**Reference conditions:** An air factor of $\lambda$ = 1.17, which corresponds to the reference case, and an ambient temperature of 8.5 °C, which is the long-term average annual temperature, were selected as reference conditions for the simulation.

## 7. RESULTS AND DISCUSSION

First, a comparison was made between coal and iron operation for reference conditions ($\lambda$ = 1.17 and an ambient temperature of 8.5 °C). All simulations are based on a premixed iron-air system. Table 4 summarizes the main results of the simulation. The efficiencies are defined in a conventional manner, as reported in Table 1. Table 4 lists the most relevant boiler temperatures: the adiabatic combustion temperature, together with the temperatures at the end of the burner level, at the end of the combustor, after the last high-pressure superheater, after the last intermediate-pressure superheater, after the economizer (ECO), and after the air preheater. These positions are highlighted in Figure 4.

Table 4 – Efficiencies and characteristic boiler temperatures for coal and iron operation.

| Efficiencies and flue gas temperatures | Position in Fig. 4 | Unit | Coal operation | Fe operation | Fe operation with Biflux heat exchanger |
|---|---|---|---|---|---|
| Total efficiency $\eta_t$ and Δ $\eta_t$ | | % | 44.1 | +1.3 | +1.5 |
| Cycle efficiency $\eta_c$ and Δ $\eta_c$ | | % | 50.8 | +-0 | +0.1 |
| Auxiliary efficiency $\eta_a$ and Δ $\eta_a$ | | % | 92.4 | +1.0 | +1.2 |
| Boiler efficiency $\eta_b$ and Δ $\eta_b$ | | % | 94.0 | +1.7 | +1.7 |
| Adiabatic flame temperature | 1 | °C | 1991 | 1961 | 1961 |
| Flue gas temperature after burner level $T_b$ | 2 | °C | 1506 | 1552 | 1551 |
| Flue gas temperature after combustor $T_c$ | 3 | °C | 1322 | 1288 | 1287 |
| Flue gas temperature after hp superheater 3 $T_{hps3}$ | 4 | °C | 1066 | 932 | 939 |
| Flue gas temperature after ip superheater 2 $T_{hps3}$ | 5 | °C | 943 | 731 | 744 |
| Flue gas temperature after ECO $T_{eco}$ | 6 | °C | 416 | 347 | 348 |
| Air temperature after air preheater $T_{aph}$ | 7 | °C | 142 | 118 | 118 |

The main findings from this simulation can be summarized as follows:

**1)** The adiabatic combustion temperatures differ by approx. 30 °C, essentially due to the 30 °C higher preheating during coal operation.

**2)** At the outlet of the boiler, the temperatures are slightly lower for iron operation due to the more intense solid radiation.

**3)** The temperatures in the superheater sections particularly reflect the significantly higher heat transfer for iron operation, due to the higher solid radiation. For reference, the heat transfer coefficients are listed in

**4)** Table 5 for two heat exchangers: the last superheaters in the high-pressure (hp) and intermediate-pressure (ip) sections, respectively. The convective heat transfer is slightly lower for iron operation due to the lower velocities. However, the radiative heat transfer is about 3 times greater due to the higher particle density.

Table 5 – Heat transfer coefficients for convection and radiation in selected heat exchangers.

| Component and heat transfer coefficient (HTC) | Units | Operation |
|---|---|---|

[2] False air (also air leakage or sucked-in air) refers to air that is added to a controlled flow of air in an uncontrolled manner, so that there is more air than specified or previously measured.

| | | Coal | Fe |
|---|---|---|---|
| HTC convection $\alpha_c$ in hp superheater 3 | W/m²K | 53 | 46 |
| HTC radiation $\alpha_r$ in hp superheater 3 | W/m²K | 79 | 242 |
| HTC convection $\alpha_c$ in ip superheater 2 | W/m²K | 57 | 248 |
| HTC radiation $\alpha_r$ in ip superheater 2 | W/m²K | 39 | 131 |

**5)** This situation is present in all heat exchangers in the convection section, meaning that the exhaust gas temperatures downstream of the ECO and downstream of the air preheater are significantly lower. This leads to an improved boiler efficiency of more than one percent.

**6)** Obviously, there is also a lower auxiliary demand since mills, the desulfurization and the denitrification can all be omitted. Although dust collection is more costly due to the high-pressure losses, a better auxiliary efficiency of more than one percent remains.

**7)** The pure cycle efficiency is expected to be about 0.1% higher for iron operation, since the steam preheater is not required. However, the greater heat transfer in the high-pressure section causes a reduction in efficiency due to the skew in the boiler, higher transferred heat flows in the high-pressure section and lower heat flows in the intermediate-pressure section. This can again be compensated for by a Biflux heat exchanger. It can be seen from Table 4 that this measure results in an efficiency improvement of almost exactly the theoretical 0.1 percent.

The reference simulations with iron, which are listed in Table 4, include several uncertainties that can be evaluated through parameter variations and systematic sensitivity analysis.

Table 6 – Plant efficiencies as a function of different operating conditions. The iron operating points S2–S10 utilize a Biflux heat exchanger. S1 and S2: see Table 2. S3: Fe operation with $\lambda = 1.4$. S4: Fe operation with $\lambda = 1.6$ and staged combustion. S5: Fe operation with $\lambda = 1.95$. S6: Fe operation with $\lambda = 1.17$ and reduced heat surface fouling. S7: Fe operation with $\lambda = 1.17$ and convective heat transfer doubled. S8: Fe operation with $\lambda = 1.17$ and convective heat transfer quadrupled. S9: Fe operation with $\lambda = 1.17$ and oxidation to $Fe_3O_4$. S10: Fe operation with $\lambda = 1.17$ and optimized fuel feeding system.

| Fuel | Coal | Fe | Fe | Fe | Fe | Fe | Fe | Fe | Fe | Fe |
|---|---|---|---|---|---|---|---|---|---|---|
| Case identifier | S1 | S2 | S3 | S4 | S5 | S6 | S7 | S8 | S9 | S10 |
| Efficiency total $\eta_t$ and $\Delta\ \eta_t$ (in %) | **44.1** | **+1.5** | **+1.2** | **+0.9** | **+0.0** | **+1.5** | **+1.7** | **+1.8** | **+1.4** | **+2.3** |
| Efficiency cycles $\eta_c$ and $\Delta\ \eta_c$ (in %) | 50.8 | +0.1 | +0.1 | +0.1 | +0.1 | +0.1 | +0.1 | +0.1 | +0.1 | +0.1 |
| Efficiency auxiliary $\eta_a$ and $\Delta\ \eta_a$ (in %) | 92.4 | +1.2 | +1.1 | +0.9 | +0.4 | +1.2 | +1.2 | +1.2 | +1.1 | +1.1 |
| Efficiency boiler $\eta_b$ and $\Delta\ \eta_b$ (in %) | 94.0 | +1.7 | +1.2 | +0.6 | -0.7 | +1.7 | +2.1 | +2.3 | +1.5 | +3.4 |

Table 6 lists a total of ten of these simulation cases. Cases S1 and S2 correspond to the two cases already presented in Table 4 and are again included in Table 6 for reference. Simulations S3 to S5 involve varying the air factor $\lambda$ with the aim of demonstrating how a lower adiabatic combustion temperature affects efficiency. As already described, particle agglomeration must always be avoided. Whether this can be achieved with an air factor as low as $\lambda = 1.17$ is a question that remains to be answered. An air factor of $\lambda = 1.95$ (see simulation S5) enables an adiabatic combustion temperature below the melting point of $Fe_3O_4$ in all cases. The efficiency deteriorates by 1.4% compared to reference case S1, but it is still comparable to the coal mode. Moreover, simulation S4 with $\lambda = 1.6$ and staged combustion shows very promising results. In the lower stage, 75% of the fuel is burned with $\lambda = 1.95$ and in the second stage the remaining fuel is supplied with the necessary conveying air. For this case, the temperatures always remain below 1597 °C. Notably, the efficiency is about 0.9 % better than for coal operation. The calculation with an intermediate air factor $\lambda = 1.4$ (case S3) shows consistent results between reference case S2 and case S4. The exact design of the staged combustion will be investigated in further detail in future work. In all cases, the calculation shows that operation at sufficiently low temperatures is possible while retaining a slight efficiency gain compared to coal firing.

Another unresolved question is the extent to which the efficiency of the overall plant deteriorates if the reaction of $Fe_3O_4$ to $Fe_2O_3$ remains incomplete at temperatures around 1300 °C. In case S9, it was assumed that iron oxidizes completely to produce $Fe_3O_4$ and the reaction producing $Fe_2O_3$ is omitted. Of course, in this case the lower calorific value of $Fe_3O_4$, approx. 6650 kJ/kg, must be used, since exactly this difference in calorific value is also saved in the reduction step that is not considered here. The efficiency deteriorates only very slightly based on this assumption.

In case S10, the retrofit is calculated with the assumption that the heat transfer surfaces in the air preheater are enlarged and that the iron fuel is preheated indirectly by the flue gas via a heat displacement system. The improved boiler efficiency (the flue gas temperatures here are around 70 °C) increases the efficiency of the overall process by 2.3%. A detailed analysis will show whether an investment of this kind is economically viable.

When building a new plant, further improvements are conceivable. These include using the latest turbine generation and reducing condensation in the preheaters and the condenser. Efficiencies of 47–48% can then be expected for power plants in this 600 °C class. If the steam parameters of the power plant were further increased to 750 °C, for example, the net efficiency potential would increase to over 50%.

## 8. SUMMARY AND CONCLUSIONS

This paper estimated the retrofit potential of an existing coal-fired power plant for operation with green iron as a fuel. This analysis is based on a case study of a modern coal-fired power plant in the 800-MW class with high steam parameters. A simulation model that had been adapted to real power plant operating data with coal firing was extended to cover iron operation utilizing EBSILON® Professional. The adaptation of the model required new components for a number of power plant sub-processes, as well as the integration of thermochemical properties for iron and its oxides. The central findings of this work can be summarized as follows:

To retrofit an existing coal power plant for iron operation, the fuel feeding system, the burners and the dedusting system have to be redesigned. Depending on the operating conditions selected, a Biflux heat exchanger or a change in the piping may still be necessary, since heat shifts can occur between the high-pressure and intermediate-pressure sections. These have a negative effect on the thermodynamic efficiency.

The retrofitted power plant achieves efficiencies that are around 1–2% higher than the efficiencies achieved with coal-firing. The main reasons for this efficiency gain are the lower internal energy consumption from auxiliary systems, since the mill and the desulfurization systems are not required, and it is very likely that the denitrification system can also be omitted. Furthermore, the heat transfer coefficients are significantly higher due to the high particle radiation, such that the exhaust gas temperature drops, and the boiler efficiency increases. Additionally, restrictions such as sulfur dew point undershoots in the flue gas segment or compliance with maximum temperatures at the mills in the air segment do not have to be satisfied by the design of the air-flue gas system. Therefore, higher air preheating temperature is feasible.

In order to leverage these attractive potentials, a number of scientific questions have to be clarified. These include:

**1)** The kinetics of the iron-air system in the high-temperature range between 1500 °C and 2000 °C and in the medium-temperature range of approximately 1300 °C;

**2)** The dynamics of heavily particle-laden three-phase flows, such as turbulence-particle interaction;

**3)** The convective and radiative heat transfer of heavily particle-laden flows.

The study clearly shows that the conversion of a coal-fired power plant to iron operation is not only possible but likely entails improvements in efficiency. This is another strong argument for introducing this carbon-free energy technology, which would allow existing coal power plant infrastructure to be used with moderate adaptation and retrofit measures.

## ACKNOWLEDGMENTS

This work was funded by the Hessian Ministry of Higher Education, Research, Science and the Arts – Clean Circles cluster project. The authors thank Magnus Kircher for his support in editing figures.